\documentclass[twocolumn,pra,showpacs]{revtex4}

\usepackage{graphicx}
\usepackage{rotating}
\usepackage{amsmath}
\usepackage{amsfonts}
\usepackage{amssymb}
\usepackage{enumerate}
\usepackage{longtable}
\usepackage{dcolumn}
\usepackage{bbm}

\begin{document}

\newcommand{\sss}{_\mathrm{s}}
\newcommand{\cc}{_\mathrm{c}}
\newcommand{\ff}{_\mathrm{F}}
\newcommand{\rr}{_\mathrm{R}}
\newcommand{\lL}{_\mathrm{L}}
\newcommand{\mx}{_\mathrm{max}}

\title{Interaction Quenches of Fermi Gases}

\author{G\"otz S. Uhrig}
\affiliation{Lehrstuhl f\"{u}r Theoretische Physik I,
Technische  Universit\"{a}t Dortmund,
 Otto-Hahn Stra\ss{}e 4, 44221 Dortmund, Germany}
\email{goetz.uhrig@tu-dortmund.de}

\date{\rm\today}

\begin{abstract}
It is shown that the jump in the momentum distribution of Fermi gases 
evolves smoothly for small and intermediate times once an interaction
between the fermions is suddenly switched on. 
The jump does not vanish abruptly. 
The loci in momentum space where the jumps occur are those of
the noninteracting Fermi sea. No relaxation of the Fermi surface geometry
takes place.
\end{abstract}

\pacs{05.70.Ln, 71.10.Fd, 05.30.Fk}


\maketitle


Recently, the interest in nonequilibrium quantum physics has
risen significantly. This is due to experimental and theoretical
progress in treating quantum systems 
with time dependent parameters. No exhaustive 
 overview of the field is possible but we mention  
the manipulation of atoms in optical lattices \cite{bloch08}
and time-dependent transitions between distinct
many-body states in particular \cite{grein02a} on the experimental side.
Sudden changes due to heating by ultrashort laser pulses make
time-dependent investigations also possible in solid state systems 
\cite{perfe06}.
On the theoretical side, there have been important advancements
in new techniques, e.g.,
 time-dependent density-matrix renormalization \cite{white04a,daley04}
and numerical renormalization \cite{ander05}, nonequilibrium dynamical
mean-field theory (DMFT) \cite{freer06,eckst08,eckst09} or
forward-backward continuous unitary transformations (CUTs) \cite{moeck08}.

The main issues investigated presently are (i) 
the description of nonequilibrium stationary states and (ii)
how various quenched physical systems approach 
such states or  equilibrium states. The
existence of conserved quantities can imply that equilibrium Gibbs states
are not reached. The systems may approach generalized Gibbs states
\cite{rigol07} which include the knowledge of the 
conserved quantities, see e.g.\
Ref.\ \onlinecite{eckst08}. In particular, integrable systems with 
their macroscopic number of conserved quantities play a special role
which may lead to more subtle stationary behavior \cite{ganga08}.
The precise way how the stationary states are approached
is investigated intensivelyb by analytical \cite{calab06,cazal06,barth08} and 
numerical means \cite{kolla07,manma07,rigol08}.

In the present article, we do not focus on thermalization itself
but on the regime before, i.e., for short and intermediate times after
the quench. We consider interaction
quenches of Fermi gases at zero temperature. 
We study how the ground state $|0\rangle$ (Fermi sea) of  
a noninteracting fermionic systems evolves once an interaction 
is suddenly switched on. This question has been investigated
in the continuum field theory of a Tomonaga-Luttinger (TL) model 
\cite{cazal06,iucci09} in one dimension (1D). 
In leading second order in the interaction $U$ of the Hubbard model
in $d>1$ dimension, the behavior of the momentum distribution (MD)
was elucidated by CUTs  \cite{moeck08}.
Very recently, nonequilibrium DMFT was used
to study the issue for $d=\infty$ dimensions \cite{eckst09}.

The latter studies motivate our investigation in particular.
The numerical data \cite{eckst09} has been consistently analyzed in terms of
a finite jump 
$\Delta n(t):=\lim_{k\to k\ff-}n_t(k)- \lim_{k\to k\ff+}n_t(k)>0$ 
in the MD
\begin{equation}
n_t(k) = 
\langle 0|e^{it H} 
c^\dagger_k c^{\phantom\dagger}_k e^{-it H}|0\rangle
\end{equation}
which decreases quickly. If the interaction is large enough
oscillatory behavior in $\Delta n(t)$ is found. This was also seen
in the numerical analysis of spinless 1D fermions \cite{manma07}.

In their study of prethermalization Moeckel and Kehrein also discuss
the short and intermediate time behavior and
state that the fermions at the Fermi momentum
acquire a finite life due to fourth order processes \cite{moeck08}. 
Thus they presume that the jump collapses immediately after the quench. 
The width $\Delta k$, over 
which the jump is broadened, is small of the order of $(\rho\ff U)^4$
where $\rho\ff$ is the density-of-states at the Fermi level.

In view of these contradictory results we aim at elucidating the 
behavior of the MD at small and intermediate times. We provide
evidence that the MD displays a jump without broadening. 
The quintessential reason is that 
long-range correlations of the quenched system remain determined
by the system before the quench.

First, we revisit the 1D TL model and extend Cazalilla's results 
\cite{cazal06,iucci09} to small and intermediate times. 
A finite jump occurs computed for all times in all orders of
the interaction \footnote{We assume that the interaction
can be tuned so that zero interaction is accessible.
If this is experimentally not possible no jump in the mathematical
sense will occur in 1D
because even a small interaction implies a continuous MD. But the
deviation in the sense of a ${\cal L}_2$ norm  from a real jump
tends to zero for vanishing interaction so that for practical purposes,
i.e., measurements with finite resolution, this fundamental
aspect will not dominate.}.
Recall that the scattering processes are particularly strong in 1D:
They destroy the conventional Fermi liquid.
Thus the survival of a finite MD jump in $d=1$ can be seen as evidence
for the persistence of the MD jump also in higher dimensions
although an alternative view is to take the 1D situation 
as being too special to deduce generic behavior in higher dimensions.
Hence, in order to support our view that the 1D behavior
is generic for the \emph{short and intermediate times} after a quench
 we secondly discuss the general situation in the Heisenberg picture.

\paragraph{Tomonaga-Luttinger Model}
1D fermionic models with linear dispersion and without
Umklapp scattering can be mapped to free bosonic models, see
e.g.\ Refs.\ \onlinecite{meden92,voit95,delft98}.
For simplicity, we first consider spinless fermions with the 
bosonized Hamiltonian in momentum space
\begin{equation}
\label{eq:hamilton}
H=\sum_{q\neq 0}\widetilde v |q|b^\dagger_qb^{\phantom\dagger}_q
+ \frac{1}{2\pi}\sum_{q\neq 0} U(q) |q| (b^\dagger_{-q} b^\dagger_{q}+
b^{\phantom{\dagger}}_{-q} b^{\phantom{\dagger}}_{q})
\end{equation}
where $\widetilde v$ is the bare Fermi velocity and 
$U(q)$ is essentially the Fourier transform of the 
density-density interaction \cite{uhrig04b}. 
The noninteracting Hamiltonian $H_0$ is the one with $U(q)=0$.
Contributions at $q=0$ are left out because they
do not matter in the sequel.
The bosonic creation operator is given in terms of the
fermionic creation (annihilation) operators $c^\dagger_k$ ($c_k$)
by $b^\dagger_q=i\sqrt{2\pi/(|q|L)}\sum_k 
c^\dagger_{k+q}c^{\phantom{\dagger}}_k$ where $L$ is
the length of the periodic system.
For the MD around the right Fermi point at $k\ff$ 
 we need the 1-particle correlation
\begin{equation}
\label{eq:G}
G\rr(x,t) = -i \langle 0| \hat\psi\rr(x,t)\hat\psi\rr(0,t) |0\rangle
\end{equation}
 where $\hat\psi\rr(x,t)=e^{itH}\hat\psi\rr(x)e^{-itH}$ 
is the annihilating real-space
field operator of the right-movers (subscript $\mathrm{L}$
is for left-movers) at site $x$ \cite{meden92,uhrig04b}.
Fourier transform of $G\rr(x,t)$ to momentum space provides
the wanted MD. The standard bosonization identity 
\cite{meden92,voit95,delft98,uhrig04b} reads
\begin{equation}
\label{eq:boson-id}
\hat\psi\rr(x) =e^{-i\varphi^\dagger\rr(x)}
e^{-i\varphi\rr(x)} U^-\rr e^{(2\pi i x/L)(\hat N\rr-\frac{1}{2})}/\sqrt{L},
\end{equation}
where $U^-\rr$ is the Klein factor decreasing the number 
$\hat N\rr$ of right-movers by one. The bosonic field $\varphi$
is given by
$\varphi_\alpha(x) = -\sigma_\alpha\sum_{\sigma_\alpha q>0}
\sqrt{2\pi/(|q|L)}
b_qe^{iqx-a|q|/2}$ where $\alpha\in\{\mathrm{R},\mathrm{L}\}$ 
and $\sigma\rr=1$ ($\sigma\lL=-1$ ),
Finally, the convergence factor $a$ will be sent to $0+$.

For evaluating $G\rr(x,t)$ we need the time-dependent
operator  $\hat\psi\rr(x,t)$ which is given by
\eqref{eq:boson-id} with $\varphi\rr(x)$ being replaced
by $\varphi_\alpha(x,t) = e^{itH} 
\varphi_\alpha(x)e^{-itH}$. This operator can be computed if
$H$ in \eqref{eq:hamilton} is diagonalized by the appropriate
Bogoliubov transform $b_q= C \tilde b_q + S \tilde b_{-q}^\dagger$
with $C=\cosh\theta_q$ and $S=\sinh\theta_q$. 
The diagonalized Hamiltonian 
$H=E_0+\sum_{q\neq 0} v |q|\tilde b^\dagger_q\tilde b^{\phantom\dagger}_q$
is characterized by the velocity $v$. For $q$-independent
$\theta_q$ the calculation is particularly transparent. Applying 
the Bogoliubov transform forward and backward \cite{cazal06,moeck08}
we eventually obtain
\begin{eqnarray}
\nonumber
\varphi\rr(x,t) &=& C^2 \varphi\rr(x-vt)
+SC\varphi\lL^\dagger(x-vt)\\
&&-SC\varphi\lL^\dagger(x+vt)) - S^2 \varphi\rr(x+vt).
\label{eq:varphi-result}
\end{eqnarray}
Combining this result with the time-dependent version of
\eqref{eq:boson-id} and inserting it in \eqref{eq:G}
yields the expectation value of the product of
exponentials in $\varphi^{(\dagger)}_\alpha(x\pm vt)$
and  $\varphi^{(\dagger)}_\alpha(\pm vt)$. Such a product
can be evaluated by bringing the annihilating
operators $\varphi$ to the right and the creating operators $\varphi^\dagger$
to the left with the help of the Baker-Campbell-Hausdorfff
formula $e^Ae^B=e^{A+B+[A,B]/2}$ if $[A,B]$ is a number only.
The required commutators are $[\varphi_\alpha(x),\varphi^\dagger_\alpha(x')]
=-\ln(2\pi i(\sigma_\alpha(x'-x)-ia)/L)$. The final result reads
\begin{eqnarray}
\label{eq:sl-result}
G\rr(x,t) &=& \frac{e^{ik\ff x}}{2\pi (x+ia)}
f(v,r)^{2\gamma(1+\gamma)}\\
\label{eq:f-def}
f(v,r) &=& \frac{r^2}{r^2+(2vt)^2}\frac{|(x+ir)^2-(2vt)^2|}{r^2+x^2},
\end{eqnarray}
with $\gamma:=S^2={\cal O}(U(0)^2)$. For $q$ independent interaction
one has $r$ in \eqref{eq:f-def} is given by the convergence 
factor $a$ which spoils the proper
limit $a\to 0+$. Conventionally, this is solved by taking into account
that the interaction is not completely local but has a range $r$.
In equilibrium calculations the most convenient assumption
is $\sinh^2\theta_q =\gamma e^{-r|q|}$ \cite{luthe74b,meden92}. 
In the present nonequilibrium context it is more convenient to assume
$\sinh^2\theta_q(1+\sinh^2\theta_q)=\gamma(1+\gamma)e^{-r|q|}$
which allows us to discuss Eq.\ \eqref{eq:sl-result} rigorously 
at $a= 0+$ for all times $t$ and distances $x$.
In this case, the above sketched derivation must be done
for each pair of modes $q,-q$ and the commutators
finally imply (\ref{eq:sl-result},\ref{eq:f-def}).
For comparison, we remind the reader that the equilibrium correlation 
\cite{meden92} reads $G_{\mathrm{R,eq}}(x) =\frac{e^{ik\ff x}}{2\pi (x+ia)}
(\frac{r^2}{r^2+x^2})^{\gamma}$.

\begin{figure}[ht]
    \begin{center}
    \includegraphics[width=\columnwidth,clip]{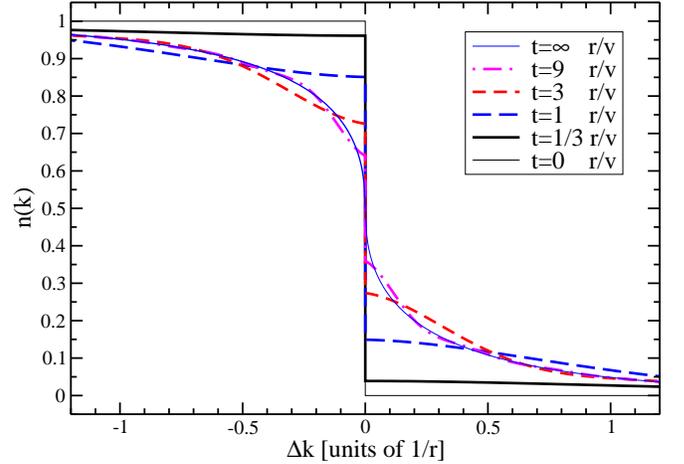}
    \end{center}
    \caption{(color online) 
      Time evolution of the momentum distribution $n_t(k)$ around the right
      Fermi point $k\ff>0$ ($\Delta k:=k-k\ff$) for the spinless TL-model 
      with $\gamma=0.1$.
      \label{fig:spinless}
}
\end{figure}
For $t=0$ in (\ref{eq:sl-result},\ref{eq:f-def}), 
$f(v,r)=1$ holds and the MD resulting from
the Fourier transform of  $e^{ik\ff x}/(2\pi (x+i0+))$ is
the expected step function of the Fermi sea, cf.\  Fig.\
\ref{fig:spinless}. For $t=\infty$
we have $f(v,r)=r^2/(r^2+x^2)$ which implies a power law
$n(k)-1/2\propto \Delta k^{4\gamma(1+\gamma)}$ without a jump.
Although a stationary correlation has been reached no thermalization
has taken place because of the simplicity of the model: macroscopic
number of conserved quantities, too little variation in the dispersion 
\cite{cazal06,barth08}.

We focus on small and intermediate times for which we find that
$G\rr(x,t)$ decreases always like $1/x$ which implies a finite
jump at the Fermi vector. The time-dependent prefactor of
$1/x$ is given by the first fraction $r^2/(r^2+(2vt)^2)$ in \eqref{eq:f-def}
to the power $2\gamma(1+\gamma)$ which agrees for $t\to\infty$
with previous analyses \cite{cazal06,iucci09}.
Hence there is a completely smooth decrease of the jump
\begin{equation}
\Delta n(t)= [r^2/(r^2+(2vt)^2)]^{2\gamma(1+\gamma)},
\end{equation}
remaining finite at all finite times. The complete MD at various
times is obtained by numerical Fourier transformation;
see Fig.\ \ref{fig:spinless}. 
In the estimate $r/v\approx \hbar/J ={\cal O}(1\mathrm{ms})$
$J$ is the hopping element from site to site which
leads to the stated time scales for
atoms in optical lattices \cite{grein02a,bloch08}.  The MD
$n_t(k)$ at given $k$ does not evolve monotonically in
time but displays oscillations in line with previous 
perturbative \cite{moeck08} and numerical \cite{manma07}
results.

 Inspecting \eqref{eq:varphi-result}
it is obvious why the $1/x$ proportionality of $G\rr(x,t)$ does not change
in the course of time. The operator $\varphi\rr(x,t)$ propagates through space
at maximum with speed $v$. Hence there is no way how for a given 
time the long distance behavior is changed. This argument relies on
the existence of a maximum speed $v\mx<\infty$ by which information can
travel through the system. This phenomenon was called light-cone effect 
by Calabrese and Cardy \cite{calab06}; 
they used the term for the propagation of 
entangled quasiparticles a quench. Note for later discussion that
such a maximum speed generally exists  independently of the
system's dimension.

But the prefactor of the
$1/x$ correlation changes in time in spite of the light-cone effect.
This represents a major change in the correlations. Inspecting our
derivation we see that this effect stems from local
commutators, i.e., from commutators between $\varphi_\alpha(x\pm vt)$
and $\varphi_\alpha^\dagger(x\mp vt)$ or from the corresponding pair at $x=0$.
Thus there is a multiplicative renormalization 
of the matrix element which links $\hat\psi^\dagger(x,t)\rr|0\rangle$
to $\hat\psi^\dagger(x)\rr|0\rangle$. Below we show
that this is also the generic situation in higher dimensions.
Here we point out that for the behavior over short distances and short times
the TL model is not  special. All corrections which might be induced by
other terms,  which are present in more general models, will not change
the qualitative behavior found here because they can be treated for
short distances and times perturbatively.

For completeness we wish to extend the results for the spinless TL
model to its counterpart with spin where charge ($\mathrm{c}$) and
spin ($\mathrm{s}$) bosonic operators arise from the 
symmetric and antisymmetric, respectively, combination of the 
$\uparrow$ and $\downarrow$ bosonic operators \cite{meden92,voit95}.
The Hamiltonian is diagonal in charge and spin bosons. 
It is characterized by the charge triple 
$v\cc,r\cc, \gamma\cc$ and the spin triple $v\sss,r\sss, \gamma\sss$.
In the extended boson identity \eqref{eq:boson-id} the sum
of the charge $\varphi_{\mathrm{R,c}}$ and the spin  $\varphi_{\mathrm{R,s}}$
modes occur multiplied by $1/\sqrt{2}$ due to normalization \cite{meden92}.
Pursuing  the same manipulations as in the spinless case eventually
leads to
\begin{equation}
\label{eq:sf-result}
G_{\mathrm{R}}(x,t) = \frac{e^{ik\ff x}}{2\pi (x+ia)}
f(v\cc,r\cc)^{\gamma\cc(1+\gamma\cc)}
f(v\sss,r\sss)^{\gamma\sss(1+\gamma\sss)}.
\end{equation}
On the one hand, many qualitative features are the same as in the spinless
case, in particular the persistence of a finite jump in the MD for all
times. It is given by
\begin{equation}
\Delta n(t)= 
\Big[\frac{r\cc^2}{r\cc^2+(2v\cc t)^2}\Big]^{\gamma\cc(1+\gamma\cc)}
\Big[\frac{r\sss^2}{r\sss^2+(2v\sss t)^2}\Big]^{\gamma\sss(1+\gamma\sss)}.
\end{equation}
On the other hand, the appearance of two velocities, two  length
scales and two exponents which are possibly different leads to
a richer phenomenology. 
We refrain from showing explicit results because the MDs look
very similar to the ones in Fig.\ \ref{fig:spinless}.
The difference in the decrease with 
the distance $x$ between Eqs.\ \eqref{eq:sl-result} and \eqref{eq:sf-result}
is hardly discernible. Hence spin-charge separation,
i.e., the difference $v_\mathrm{c}\ne v_\mathrm{s}$ is
 visible only in high precision measurements.

\paragraph{General Interacting Fermions}
Because the scattering
induced by interaction is particularly strong in 1D,
 the persistence of the MD jump in 1D 
indicates that it should
persist in higher dimensions as well. Alternatively, one
may presume the persistence of the jump to be 
peculiar to 1D. To support our view that the persistence
is generic we will analyze the general
equations of motion below.
Note that in all dimensions a maximum
speed $v\mx <\infty $ exists at which operators can propagate. Thus the
long range behavior of correlations cannot change within a finite
times. 

To be specific we consider the 1-particle
correlation $G(\mathbf{r},t)=
-i\langle 0| \hat\psi(\mathbf{r},t)\hat\psi^\dagger(\mathbf{0},t)
|0\rangle$ where the Heisenberg time evolution of the operators is induced by
the interacting Hamiltonian $H$ while $|0\rangle$ is the 
Fermi sea of the noninteracting Hamiltonian $H_0$. A  spin
dependence is omitted for simplicity. The Heisenberg
equation  reads
$\partial_t \hat\psi^\dagger(\mathbf{r},t)= 
i [H,\hat\psi^\dagger(\mathbf{r},t)]$.
The commutation with $H_0$ (Liouville operator
$L_0$) will propagate the 1-particle operator
only. The commutation with the interaction ($L_\mathrm{I}$)
generates a particle-hole (PH) pair. The iteration of $L_\mathrm{I}$
increases the number of particle-hole pairs. Hence the
structure of the solution is
\begin{equation}
\label{eq:structure}
\hat\psi^\dagger(\mathbf{r},t) = P^\dagger_\mathbf{r}+
P^\dagger (P^\dagger H^\dagger)_\mathbf{r} +
P^\dagger(P^\dagger H^\dagger)^2_\mathbf{r} + \ldots
\end{equation}
where $P^\dagger$ ($H^\dagger$) stands for a created particle (hole).
It is understood that each term  in \eqref{eq:structure} is
normal-ordered relative to the Fermi sea $|0\rangle$.
The application $L_0$ will reproduce the structure of a 
term $P^\dagger(P^\dagger H^\dagger)^m_\mathbf{r}$, i.e., $m$ stays
fixed, while $L_\mathrm{I}$ can increase $m$ by one, leave it unchanged, 
decrease it by one or by two. The first two terms in \eqref{eq:structure}
are denoted explicitly in $d$ dimensions
\begin{eqnarray}
&&P^\dagger_\mathbf{r} = \int_{|\mathbf{r}_1|<v\mx t}
h_0(\mathbf{r}_1,t):\hat \psi^\dagger(\mathbf{r}_1+\mathbf{r}):d^dr_1
\\
\nonumber
&&P^\dagger (P^\dagger H^\dagger)_\mathbf{r} =
\iiint_{|\mathbf{r}_j|<v\mx t}
h_1(\mathbf{r}_1,\mathbf{r}_2,\mathbf{r}_3,t)\cdot
\\
&&
:\hat \psi^\dagger(\mathbf{r}_1+\mathbf{r})
\hat \psi^\dagger(\mathbf{r}_2+\mathbf{r})
\hat \psi(\mathbf{r}_3+\mathbf{r}):d^dr_1d^dr_2d^dr_3.
\qquad
\end{eqnarray}
The Heisenberg equation generates a hierarchy of 
coupled differential equations for the $h_m(\{\mathbf{r}_j\},t)$.
Because $h_m(\{\mathbf{r}_j\},0)=0$ for $m>0$ and a term with $m$ PH pairs
requires at least $m$ applications of $L_\mathrm{I}$
we know $h_m ={\cal O}((Ut)^m)$ where $U$ is a generic value of the
interaction. The expansion in $t$ can be computed order by order.
Though we cannot prove the convergence of the series $t$ we do not see any
means that the radius of convergence vanishes for Hamiltonians with
finite coefficients. Certainly, the hierarchy is finite and thus
well-behaved if calculations up to a finite order in the interaction, 
e.g., $U^4$, are carried out.

For $G(\mathbf{r},t)$ the expectation value
$ \langle 0| \hat\psi(\mathbf{r},t)\hat\psi^\dagger(\mathbf{0},t)
|0\rangle$ must be evaluated which contains terms like
\begin{equation}
\label{eq:structure-result}
\langle 0| (H P)^m P_\mathbf{r}^{\phantom j}
P^\dagger(P^\dagger H^\dagger)^j_\mathbf{0}\rangle
=\delta_{m,j} G^{(2m+1)}(\mathbf{r},t)
\end{equation}
because only states with the same number
of PH pairs can have a finite overlap. The Wick theorem is
applicable because $|0\rangle$ is a Fermi sea so that the many-particle
correlations can be reduced to products of the initial 1-particle correlations
$g(\mathbf{r}):=G(\mathbf{r},0)$. Thus we know $G^{(2m+1)}(\mathbf{r},t)
={\cal O}\left( g(\mathbf{r})^{2m+1}\right)$.
 The information about the jump is encoded 
in the most slowly decreasing contribution for $|\mathbf{r}|\to\infty$,
namely the one for $m=0$
\begin{eqnarray}
\nonumber
G^{(1)}(\mathbf{r},t)
&=&
\iint_{|\mathbf{r}_j|<v\mx t} 
h_0^*(\mathbf{r}_1,t) \cdot\\
&&g(\mathbf{r}+\mathbf{r}_1-\mathbf{r}_2)
h_0(\mathbf{r}_2,t)d^dr_1 d^dr_2.\qquad
\end{eqnarray}
This double convolution implies
$n_t^{(1)}(\mathbf{k}) =n_0(\mathbf{k})|h_0(\mathbf{k},t)|^2$ 
in momentum space
where $n_0(\mathbf{k})\in \{0,1\}$ is the noninteracting
MD. Clearly, the nonequilibrium
1-particle correlations inherit many of their qualitative properties 
 from the 1-particle correlations before the quench. Interestingly,
the jumps in the MD occur at the
same loci where the noninteracting MD $n_0(\mathbf{k})$ jumps, i.e.,
at the noninteracting Fermi surface FS$_0$.
The reduction of the jump is given by
\begin{equation}
\Delta n(t)\big|_{\mathbf{k}\in\mathrm{FS}_0}
=|h_0(\mathbf{k},t)|^2\big|_{\mathbf{k}\in\mathrm{FS}_0}.
\end{equation}
The Fourier transform $h_0(\mathbf{k},t)$ corresponds
exactly to $h_{k\uparrow}$ in Ref.\ \onlinecite{moeck08} 
after the forward-backward CUT. 
These general equations set the stage for the analysis
of higher dimensional systems and allow us to
draw general conclusions.  Further analysis has
to be numerical and it is therefore left to future research.

\paragraph{Conclusions}
Summarizing, we studied interaction quenches of noninteracting Fermi gases.
The focus was the question how the jump in the momentum distribution (MD)
vanishes. In 1D the Tomonaga-Luttinger model was investigated quantitatively
and in higher dimension the general equations of motions were set up. 
For generic Hamiltonians without diverging coefficients we showed
that the jump survives for small and intermediate times, displaying
a smooth behavior as function of time.
If thermalization takes place, we expect that the jump decreases  exponentially
(with or without oscillations).

We found that the quenched MD still displays many qualitative features
of the noninteracting Fermi sea. In particular, the loci of the jumps
are those of the Fermi sea. The Fermi surface of
isolated quenched interacting fermion models does not evolve at all.
The models have to be extended in order to incorporate relaxation of the 
Fermi surface geometry.
Finally, we point out that the approach used here for the
general situation can also 
be extended to correlations of two or more particles.

We gratefully acknowledge many helpful discussions with F.B.\ Anders and 
S.\ Kehrein.


\begin{thebibliography}{24}
\expandafter\ifx\csname natexlab\endcsname\relax\def\natexlab#1{#1}\fi
\expandafter\ifx\csname bibnamefont\endcsname\relax
  \def\bibnamefont#1{#1}\fi
\expandafter\ifx\csname bibfnamefont\endcsname\relax
  \def\bibfnamefont#1{#1}\fi
\expandafter\ifx\csname citenamefont\endcsname\relax
  \def\citenamefont#1{#1}\fi
\expandafter\ifx\csname url\endcsname\relax
  \def\url#1{\texttt{#1}}\fi
\expandafter\ifx\csname urlprefix\endcsname\relax\def\urlprefix{URL }\fi
\providecommand{\bibinfo}[2]{#2}
\providecommand{\eprint}[2][]{\url{#2}}

\bibitem[{\citenamefont{Bloch et~al.}(2008)\citenamefont{Bloch, Dalibard, and
  Zwerger}}]{bloch08}
\bibinfo{author}{\bibfnamefont{I.}~\bibnamefont{Bloch}},
  \bibinfo{author}{\bibfnamefont{J.}~\bibnamefont{Dalibard}}, \bibnamefont{and}
  \bibinfo{author}{\bibfnamefont{W.}~\bibnamefont{Zwerger}},
  \bibinfo{journal}{Rev. Mod. Phys.} \textbf{\bibinfo{volume}{80}},
  \bibinfo{pages}{885} (\bibinfo{year}{2008}).

\bibitem[{\citenamefont{Greiner et~al.}(2002)\citenamefont{Greiner, Mandel,
  H\"ansch, and Bloch}}]{grein02a}
\bibinfo{author}{\bibfnamefont{M.}~\bibnamefont{Greiner}},
  \bibinfo{author}{\bibfnamefont{O.}~\bibnamefont{Mandel}},
  \bibinfo{author}{\bibfnamefont{T.~W.} \bibnamefont{H\"ansch}},
  \bibnamefont{and} \bibinfo{author}{\bibfnamefont{I.}~\bibnamefont{Bloch}},
  \bibinfo{journal}{Nature} \textbf{\bibinfo{volume}{415}}, \bibinfo{pages}{39}
  (\bibinfo{year}{2002}).

\bibitem[{\citenamefont{Perfetti et~al.}(2006)\citenamefont{Perfetti, Loukakos,
  Lisowski, Bovensiepen, Berger, Biermann, Cornaglia, Georges, and
  Wolf}}]{perfe06}
\bibinfo{author}{\bibfnamefont{L.}~\bibnamefont{Perfetti}},
  \bibinfo{author}{\bibfnamefont{P.~A.} \bibnamefont{Loukakos}},
  \bibinfo{author}{\bibfnamefont{M.}~\bibnamefont{Lisowski}},
  \bibinfo{author}{\bibfnamefont{U.}~\bibnamefont{Bovensiepen}},
  \bibinfo{author}{\bibfnamefont{H.}~\bibnamefont{Berger}},
  \bibinfo{author}{\bibfnamefont{S.}~\bibnamefont{Biermann}},
  \bibinfo{author}{\bibfnamefont{P.~S.} \bibnamefont{Cornaglia}},
  \bibinfo{author}{\bibfnamefont{A.}~\bibnamefont{Georges}}, \bibnamefont{and}
  \bibinfo{author}{\bibfnamefont{M.}~\bibnamefont{Wolf}},
  \bibinfo{journal}{Phys. Rev. Lett.} \textbf{\bibinfo{volume}{97}},
  \bibinfo{pages}{067402} (\bibinfo{year}{2006}).

\bibitem[{\citenamefont{White and Feiguin}(2004)}]{white04a}
\bibinfo{author}{\bibfnamefont{S.~R.} \bibnamefont{White}} \bibnamefont{and}
  \bibinfo{author}{\bibfnamefont{A.~E.} \bibnamefont{Feiguin}},
  \bibinfo{journal}{Phys. Rev. Lett.} \textbf{\bibinfo{volume}{93}},
  \bibinfo{pages}{076401} (\bibinfo{year}{2004}).

\bibitem[{\citenamefont{Daley et~al.}(2004)\citenamefont{Daley, Kollath,
  Schollw\"ock, and Vidal}}]{daley04}
\bibinfo{author}{\bibfnamefont{A.~J.} \bibnamefont{Daley}},
  \bibinfo{author}{\bibfnamefont{C.}~\bibnamefont{Kollath}},
  \bibinfo{author}{\bibfnamefont{U.}~\bibnamefont{Schollw\"ock}},
  \bibnamefont{and} \bibinfo{author}{\bibfnamefont{G.}~\bibnamefont{Vidal}},
  \bibinfo{journal}{J. Stat. Mech.: Theor. Exp.} p. \bibinfo{pages}{P04005}
  (\bibinfo{year}{2004}).

\bibitem[{\citenamefont{Anders and Schiller}(2005)}]{ander05}
\bibinfo{author}{\bibfnamefont{F.~B.} \bibnamefont{Anders}} \bibnamefont{and}
  \bibinfo{author}{\bibfnamefont{A.}~\bibnamefont{Schiller}},
  \bibinfo{journal}{Phys. Rev. Lett.} \textbf{\bibinfo{volume}{95}},
  \bibinfo{pages}{196801} (\bibinfo{year}{2005});
  \bibinfo{journal}{Phys. Rev. B} \textbf{\bibinfo{volume}{74}},
  \bibinfo{pages}{245113} (\bibinfo{year}{2006}).
  

\bibitem[{\citenamefont{Freericks et~al.}(2006)\citenamefont{Freericks,
  Turkowski, and Zlati\'c}}]{freer06}
\bibinfo{author}{\bibfnamefont{J.~K.} \bibnamefont{Freericks}},
  \bibinfo{author}{\bibfnamefont{V.~M.} \bibnamefont{Turkowski}},
  \bibnamefont{and} \bibinfo{author}{\bibfnamefont{V.}~\bibnamefont{Zlati\'c}},
  \bibinfo{journal}{Phys. Rev. Lett.} \textbf{\bibinfo{volume}{97}},
  \bibinfo{pages}{266408} (\bibinfo{year}{2006}).

\bibitem[{\citenamefont{Eckstein and Kollar}(2008)}]{eckst08}
\bibinfo{author}{\bibfnamefont{M.}~\bibnamefont{Eckstein}} \bibnamefont{and}
  \bibinfo{author}{\bibfnamefont{M.}~\bibnamefont{Kollar}},
  \bibinfo{journal}{Phys. Rev. Lett.} \textbf{\bibinfo{volume}{100}},
  \bibinfo{pages}{120404} (\bibinfo{year}{2008}).

\bibitem[{\citenamefont{Eckstein et~al.}(2009)\citenamefont{Eckstein, Kollar,
  and Werner}}]{eckst09}
\bibinfo{author}{\bibfnamefont{M.}~\bibnamefont{Eckstein}},
  \bibinfo{author}{\bibfnamefont{M.}~\bibnamefont{Kollar}}, \bibnamefont{and}
  \bibinfo{author}{\bibfnamefont{P.}~\bibnamefont{Werner}},
  \bibinfo{journal}{Phys. Rev. Lett.} \textbf{\bibinfo{volume}{103}},
  \bibinfo{pages}{056403} (\bibinfo{year}{2009}).

\bibitem[{\citenamefont{Moeckel and Kehrein}(2008)}]{moeck08}
\bibinfo{author}{\bibfnamefont{M.}~\bibnamefont{Moeckel}} \bibnamefont{and}
  \bibinfo{author}{\bibfnamefont{S.}~\bibnamefont{Kehrein}},
  \bibinfo{journal}{Phys. Rev. Lett.} \textbf{\bibinfo{volume}{100}},
  \bibinfo{pages}{175702} (\bibinfo{year}{2008});
  \bibinfo{journal}{Ann. Phys.} \textbf{\bibinfo{volume}{324}},
  \bibinfo{pages}{2146} (\bibinfo{year}{2009}).

\bibitem[{\citenamefont{Rigol et~al.}(2007)\citenamefont{Rigol, Dunjko,
  Yurovsky, and Olshanii}}]{rigol07}
\bibinfo{author}{\bibfnamefont{M.}~\bibnamefont{Rigol}},
  \bibinfo{author}{\bibfnamefont{V.}~\bibnamefont{Dunjko}},
  \bibinfo{author}{\bibfnamefont{V.}~\bibnamefont{Yurovsky}}, \bibnamefont{and}
  \bibinfo{author}{\bibfnamefont{M.}~\bibnamefont{Olshanii}},
  \bibinfo{journal}{Phys. Rev. Lett.} \textbf{\bibinfo{volume}{98}},
  \bibinfo{pages}{050405} (\bibinfo{year}{2007}).

\bibitem[{\citenamefont{Gangardt and Pustilnik}(2008)}]{ganga08}
\bibinfo{author}{\bibfnamefont{D.~M.} \bibnamefont{Gangardt}} \bibnamefont{and}
  \bibinfo{author}{\bibfnamefont{M.}~\bibnamefont{Pustilnik}},
  \bibinfo{journal}{Phys. Rev. A} \textbf{\bibinfo{volume}{77}},
  \bibinfo{pages}{041604(R)} (\bibinfo{year}{2008}).

\bibitem[{\citenamefont{Calabrese and Cardy}(2006)}]{calab06}
\bibinfo{author}{\bibfnamefont{P.}~\bibnamefont{Calabrese}} \bibnamefont{and}
  \bibinfo{author}{\bibfnamefont{J.}~\bibnamefont{Cardy}},
  \bibinfo{journal}{Phys. Rev. Lett.} \textbf{\bibinfo{volume}{96}},
  \bibinfo{pages}{136801} (\bibinfo{year}{2006}).

\bibitem[{\citenamefont{Cazalilla}(2006)}]{cazal06}
\bibinfo{author}{\bibfnamefont{M.~A.} \bibnamefont{Cazalilla}},
  \bibinfo{journal}{Phys. Rev. Lett.} \textbf{\bibinfo{volume}{97}},
  \bibinfo{pages}{156403} (\bibinfo{year}{2006}).

\bibitem[{\citenamefont{Barthel and Schollw\"ock}(2008)}]{barth08}
\bibinfo{author}{\bibfnamefont{T.}~\bibnamefont{Barthel}} \bibnamefont{and}
  \bibinfo{author}{\bibfnamefont{U.}~\bibnamefont{Schollw\"ock}},
  \bibinfo{journal}{Phys. Rev. Lett.} \textbf{\bibinfo{volume}{100}},
  \bibinfo{pages}{100601} (\bibinfo{year}{2008}).

\bibitem[{\citenamefont{Kollath et~al.}(2007)\citenamefont{Kollath, L\"auchli,
  and Altman}}]{kolla07}
\bibinfo{author}{\bibfnamefont{C.}~\bibnamefont{Kollath}},
  \bibinfo{author}{\bibfnamefont{A.~M.} \bibnamefont{L\"auchli}},
  \bibnamefont{and} \bibinfo{author}{\bibfnamefont{E.}~\bibnamefont{Altman}},
  \bibinfo{journal}{Phys. Rev. Lett.} \textbf{\bibinfo{volume}{98}},
  \bibinfo{pages}{180601} (\bibinfo{year}{2007}).

\bibitem[{\citenamefont{Manmana et~al.}(2007)\citenamefont{Manmana, Wessel,
  Noack, and Muramatsu}}]{manma07}
\bibinfo{author}{\bibfnamefont{S.~R.} \bibnamefont{Manmana}},
  \bibinfo{author}{\bibfnamefont{S.}~\bibnamefont{Wessel}},
  \bibinfo{author}{\bibfnamefont{R.~M.} \bibnamefont{Noack}}, \bibnamefont{and}
  \bibinfo{author}{\bibfnamefont{A.}~\bibnamefont{Muramatsu}},
  \bibinfo{journal}{Phys. Rev. Lett.} \textbf{\bibinfo{volume}{98}},
  \bibinfo{pages}{210405} (\bibinfo{year}{2007}).

\bibitem[{\citenamefont{Rigol et~al.}(2008)\citenamefont{Rigol, Dunjko, and
  Olshanii}}]{rigol08}
\bibinfo{author}{\bibfnamefont{M.}~\bibnamefont{Rigol}},
  \bibinfo{author}{\bibfnamefont{V.}~\bibnamefont{Dunjko}}, \bibnamefont{and}
  \bibinfo{author}{\bibfnamefont{M.}~\bibnamefont{Olshanii}},
  \bibinfo{journal}{Nature} \textbf{\bibinfo{volume}{452}},
  \bibinfo{pages}{854} (\bibinfo{year}{2008}).

\bibitem[{\citenamefont{Iucci and Cazalilla}(2009)}]{iucci09}
\bibinfo{author}{\bibfnamefont{A.}~\bibnamefont{Iucci}} \bibnamefont{and}
  \bibinfo{author}{\bibfnamefont{M.~A.} \bibnamefont{Cazalilla}},
  \bibinfo{journal}{arXiv:0903.1205}  (\bibinfo{year}{2009}).

\bibitem[{\citenamefont{Meden and Sch\"onhammer}(1992)}]{meden92}
\bibinfo{author}{\bibfnamefont{V.}~\bibnamefont{Meden}} \bibnamefont{and}
  \bibinfo{author}{\bibfnamefont{K.}~\bibnamefont{Sch\"onhammer}},
  \bibinfo{journal}{Phys. Rev. B} \textbf{\bibinfo{volume}{46}},
  \bibinfo{pages}{15753} (\bibinfo{year}{1992}).

\bibitem[{\citenamefont{Voit}(1995)}]{voit95}
\bibinfo{author}{\bibfnamefont{J.}~\bibnamefont{Voit}}, \bibinfo{journal}{Rep.
  Prog. Phys.} \textbf{\bibinfo{volume}{58}}, \bibinfo{pages}{977}
  (\bibinfo{year}{1995}).

\bibitem[{\citenamefont{von Delft and Schoeller}(1998)}]{delft98}
\bibinfo{author}{\bibfnamefont{J.}~\bibnamefont{von Delft}} \bibnamefont{and}
  \bibinfo{author}{\bibfnamefont{H.}~\bibnamefont{Schoeller}},
  \bibinfo{journal}{Ann. Physik} \textbf{\bibinfo{volume}{7}},
  \bibinfo{pages}{225} (\bibinfo{year}{1998}).

\bibitem[{\citenamefont{Uhrig}(2004)}]{uhrig04b}
\bibinfo{author}{\bibfnamefont{G.~S.} \bibnamefont{Uhrig}},
\bibinfo{howpublished}{Lecture Notes available at
  \url{t1.physik.tu-dortmund.de/uhrig/liquids_ss2004.html}}.

\bibitem[{\citenamefont{Luther and Peschel}(1974)}]{luthe74b}
\bibinfo{author}{\bibfnamefont{A.}~\bibnamefont{Luther}} \bibnamefont{and}
  \bibinfo{author}{\bibfnamefont{I.}~\bibnamefont{Peschel}},
  \bibinfo{journal}{Phys. Rev. B} \textbf{\bibinfo{volume}{9}},
  \bibinfo{pages}{2911} (\bibinfo{year}{1974}).

\end{thebibliography}

\end{document}